\documentclass[11pt,letterpaper]{article}
\usepackage[margin=1in]{geometry}
\usepackage[T1]{fontenc}
\usepackage[utf8]{inputenc}
\usepackage{lmodern}
\usepackage[numbers,sort&compress]{natbib}
\usepackage{url}
\expandafter\def\expandafter\UrlBreaks\expandafter{\UrlBreaks
  \do\a\do\b\do\c\do\d\do\e\do\f\do\g\do\h\do\i\do\j\do\k\do\l\do\m
  \do\n\do\o\do\p\do\q\do\r\do\s\do\t\do\u\do\v\do\w\do\x\do\y\do\z
  \do\A\do\B\do\C\do\D\do\E\do\F\do\G\do\H\do\I\do\J\do\K\do\L\do\M
  \do\N\do\O\do\P\do\Q\do\R\do\S\do\T\do\U\do\V\do\W\do\X\do\Y\do\Z
  \do\0\do\1\do\2\do\3\do\4\do\5\do\6\do\7\do\8\do\9}
\usepackage[colorlinks=true,linkcolor=blue,citecolor=blue,urlcolor=blue]{hyperref}
\providecommand{\Description}[1]{}
\usepackage{amsmath,amsfonts,bm}

\def\eqref#1{equation~\ref{#1}}

\def\1{\bm{1}}

\DeclareMathAlphabet{\mathsfit}{\encodingdefault}{\sfdefault}{m}{sl}
\SetMathAlphabet{\mathsfit}{bold}{\encodingdefault}{\sfdefault}{bx}{n}

\newcommand{\E}{\mathbb{E}}

\newcommand{\R}{\mathbb{R}}

\usepackage{booktabs}
\usepackage{graphicx}
\usepackage{float}
\usepackage{xcolor}
\usepackage{multirow}
\usepackage{microtype}
\microtypesetup{expansion=false}
\usepackage{amsmath,amssymb}
\usepackage{algorithm}
\usepackage[noend]{algpseudocode}
\usepackage{tikz}
\usetikzlibrary{arrows.meta,backgrounds,calc,fit,positioning,shapes.geometric}

\definecolor{highblue}{HTML}{176B87}
\definecolor{loworange}{HTML}{D97706}
\definecolor{evalgreen}{HTML}{2F855A}
\definecolor{softgray}{HTML}{F3F4F6}
\definecolor{midgray}{HTML}{6B7280}
\definecolor{todored}{HTML}{B91C1C}

\newcommand{\sysname}{TileCas}

\usepackage{listings}
\usepackage{tabularx}

\definecolor{twblue}{HTML}{285A78}
\definecolor{twgray}{HTML}{F5F6F7}
\definecolor{twcomment}{HTML}{687078}

\lstdefinestyle{twcode}{
  backgroundcolor=\color{twgray},
  basicstyle=\ttfamily\fontsize{7.5}{9.5}\selectfont,
  keywordstyle=\color{twblue}\bfseries,
  commentstyle=\color{twcomment}\itshape,
  stringstyle=\color{twblue},
  columns=fullflexible,
  keepspaces=true,
  showstringspaces=false,
  breaklines=true,
  breakatwhitespace=false,
  frame=none,
  xleftmargin=5pt,
  xrightmargin=5pt,
  framexleftmargin=5pt,
  framexrightmargin=5pt,
  aboveskip=5pt,
  belowskip=7pt,
  tabsize=2
}

\newcommand{\akobase}{AKO}
\newcommand{\kdabase}{KDA}
\newcommand{\numworkloads}{50}

\newcommand{\bestbaselineavgspeedup}{\(3.75\times\)}

\newcommand{\akosysavgspeedup}{\ensuremath{4.84\times}}

\newcommand{\akoimprovement}{29.0\%}

\newcommand{\kdaimprovement}{69.2\%}

\title{\bfseries TileCas: Cascading Kernel Development Across Programming Abstractions}
\author{%
Xinyu Xiao$^{1,2,4}$ \quad Zihe Song$^{3}$ \quad
Dezhi Ran$^{2,5}$ \quad
Wei Yang$^{1,2,4,*}$ \quad Tao Xie$^{1,2,4,5,*}$\\[0.6em]
\parbox{0.96\textwidth}{\centering\small
$^{1}$College of Computer Science and Artificial Intelligence,\\
Fudan University, Shanghai, China\\[0.2em]
$^{2}$Beijing Tongming Lake Information Technology Application\\
Innovation Center, Beijing, China\\[0.2em]
$^{3}$The University of Texas at Dallas, Richardson, TX, USA\\[0.2em]
$^{4}$Institute of Systems for Advanced Computing,\\
Fudan University, Shanghai, China\\[0.2em]
$^{5}$School of Computer Science, Peking University, Beijing, China\\[0.45em]
\texttt{xiaoxy26@m.fudan.edu.cn}, \texttt{zihe.song@utdallas.edu}\\
\texttt{dezhiran@pku.edu.cn}, \texttt{yang\_wei@fudan.edu.cn}\\
\texttt{taoxie@pku.edu.cn}\\[0.35em]
$^{*}$Co-corresponding authors.}}
\date{}

\hypersetup{
  pdftitle={TileCas: Cascading Kernel Development Across Programming Abstractions},
  pdfauthor={Xinyu Xiao, Zihe Song, Dezhi Ran, Wei Yang, Tao Xie},
  pdfsubject={Cascading kernel development across programming abstractions},
  pdfkeywords={accelerator kernels, coding agents, TileLang, Ascend C, HIP}
}

\begin{document}

\maketitle

\begin{abstract}
Programming languages and abstractions have traditionally been designed for humans, but code generation is increasingly performed by models and autonomous coding agents. The abstractions that best support human programmers may therefore not be optimal for machine-driven programming. We investigate this question through GPU and AI-accelerator kernel generation, where different abstraction levels expose complementary trade-offs: high-level domain-specific languages (DSLs) simplify search by delegating implementation details to compilers, while lower-level representations provide finer-grained control for more precise performance tuning at the cost of a larger and more complex search space.

In this paper, we present \sysname{}, a unified, pluggable interface that enables coding agents to develop kernels across multiple editable abstraction levels. Rather than committing to a single representation, \sysname{} allows an agent to transfer and refine candidates across levels through \emph{Cross-Level Diagnosis}, which maps correctness and performance feedback to actionable controls at different abstraction levels; \emph{Abstraction Scheduling}, which selects both the candidate and representation from which subsequent refinement should proceed; and a \emph{Heterogeneous Search Tree}, which preserves and revisits alternative search trajectories spanning multiple representations.

We integrate \sysname{} with two code agents \akobase{} and \kdabase{}, and evaluate both integrations on \numworkloads{} end-to-end KernelBench Level~3 workloads across two accelerator platforms and two language models.
Giving equal weight to the four accelerator--model configurations, \akobase{} with \sysname{} achieves an arithmetic-mean speedup of \akosysavgspeedup{}, compared with \bestbaselineavgspeedup{} for the strongest evaluated baseline without \sysname{}, an improvement of
\akoimprovement{}.
Integrating the same interface with \kdabase{} improves its mean speedup by \kdaimprovement{}, with gains in every configuration.
These results show that allowing coding agents to navigate across programming abstractions can improve machine-driven software development and can be reused across different base workflows.

\end{abstract}


\section{Introduction}
\label{sec:introduction}

As coding agents take on more code generation tasks, understanding how programming abstractions shape code development becomes increasingly important.
Google reports that 75\% of its new code is generated by artificial intelligence (AI) systems and approved by engineers~\cite{google2026cloudnext}.
Anthropic reports that, as of May 2026, more than 80\% of its merged code was authored by Claude~\cite{anthropic2026recursive}.
This shift raises a fundamental question: are abstractions designed for human programmers also the most effective interfaces for coding agents? Because agents develop programs through repeated generation, evaluation, and refinement, the choice of abstraction directly shapes the search space that the agents and models must navigate.

Accelerator kernel programming provides a particularly useful setting for studying this question because performance depends on decisions spanning multiple abstraction levels.
High-level domain-specific languages~(DSLs), such as Triton~\cite{tillet2019triton} and TileLang~\cite{wang2025tilelang}, provide compact operations for computation and data movement while delegating some implementation details to the compiler.
In contrast, low-level native source code exposes more direct control over memory accesses, synchronization, and hardware operations, enabling optimizations unavailable at higher levels but requiring the agent to reason about a substantially larger set of interacting implementation choices.

\begin{figure}[t]
    \centering
    \includegraphics[width=0.8\linewidth]{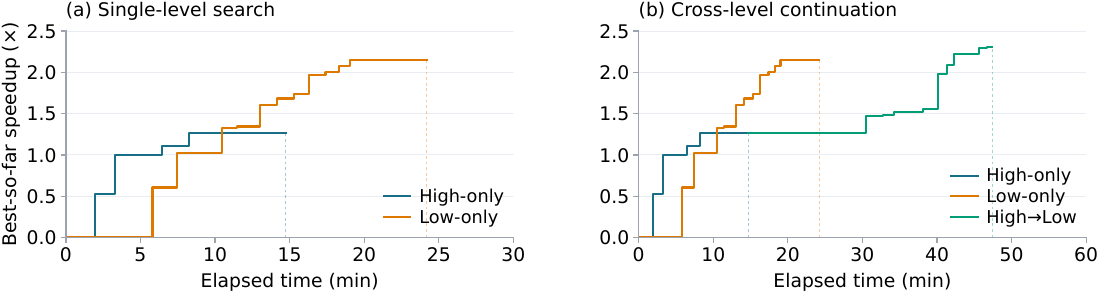}
    \caption{Performance trajectories for MinGPT on an AMD Radeon AI PRO R9700.
    High and Low denote TileLang and HIP C++, respectively.
    (a) High-only search achieves larger speedups early in the search,
    while Low-only search reaches a higher speedup later.
    (b) Continuing development at the low level from an exported
    high-level candidate yields further improvement.}
    \Description{Motivating example showing single-level search and cross-level continuation for MinGPT on an AMD accelerator.}
    \label{fig:level-compare}
\end{figure}

These properties make different abstraction levels useful at different stages of machine-driven software development. 
Early in the search, a high-level representation can help an agent rapidly explore algorithmic structures and reach a strong implementation without managing every hardware detail. Once progress at that level saturates, however, further improvement may require controls exposed only by a lower-level representation. 
Figure~\ref{fig:level-compare}(a) illustrates  this behavior: 
High-only improves more rapidly early in the search, whereas Low-only progresses slowly but with much greater cost and will eventually reach a higher performance level.
By transferring a promising high-level candidate into editable low-level code, an agent can preserve the progress already made while opening a new space of optimization opportunities;
Figure~\ref{fig:level-compare}(b) shows the further improvement obtained through continued low-level development in the same example.

This observation motivates \emph{cross-level programming}: rather than committing to a single representation throughout development, an agent should be able to move across abstraction levels as the search evolves. Existing agentic kernel workflows~\cite{ako2026,kda2026} provide the loop of candidate generation, evaluation, profiling, and iterative refinement; we refer to this underlying process as the \emph{base workflow}. Cross-level programming extends this process by allowing refinement to continue in different representations, each exposing different controls and search spaces. The key challenge is therefore not merely supporting multiple languages, but coordinating when and how the search should move across abstraction levels while retaining useful information from prior exploration.

Such coordination requires three related decisions: \emph{what} should be changed (\textit{i.e.,} identifying editable controls relevant to the feedback),  \emph{where}  the change should be made (\textit{i.e.,} selecting a level for the proposed modification), and \emph{which prior candidate} should serve as the starting point (to preserve the context of previous observations).
First, hardware evidence, such as execution timings and available profiling data, identifies observed performance behavior, but does not directly reveal the programming decisions responsible for the behavior.
A low-level bottleneck may originate from high-level choices in data layout or computation structure.
The agent must therefore relate hardware evidence back to editable controls in the program.
Second, once a modification goal is identified, the same modification goal may be realizable at multiple levels, with different degrees of control, implementation complexity, and correctness risks.
The agent must determine which level provides the most appropriate interface for the next refinement.
Third, these assessments also depend on the history of each candidate, because edits and representation conversions can change the conditions under which previous observations apply.
Cross-level programming must therefore preserve candidate histories so that each observation remains associated with the context in which it was obtained.

To address these challenges, we present \sysname{}, a pluggable method for feedback-guided programming of accelerator kernels across abstraction levels.
Given candidate code and evaluation feedback from the base workflow, \sysname{} compares development options across abstraction levels using the available evidence and candidate history.
\sysname{} returns the selected action, source candidate, and target abstraction level to guide the base workflow's next code generation or modification step.
We propose three cooperating components to support \sysname{}, as shown in Figure~\ref{fig:overview}.
\textcircled{1} \emph{Cross-Level Diagnosis} establishes source correspondence between a candidate's code representations.
Profile association links available hardware evidence to the executed code of the corresponding candidate, helping the next component examine the implementation choices behind the observed behavior and identify relevant editable controls.
\textcircled{2} \emph{Abstraction Scheduling} uses the evidence to develop modification plans, compare available controls and implementation requirements across levels, and combine these comparisons with candidate history to select a source candidate and target abstraction level.
Possible choices include further edits at the current level, a supported conversion to a lower level, and renewed exploration from a retained higher-level candidate.
\textcircled{3} \emph{Heterogeneous Search Tree} maintains the relationships between candidates created by code modifications and representation conversions, together with evaluation results for retained candidates.
These records preserve the context of each observation, helping other components assess the relevance of prior observations when selecting a candidate for further modification.

We evaluate \sysname{} using two recent agentic kernel workflows, Agentic Kernel Optimization (\akobase{})~\cite{ako2026} and Kernel Design Agents (\kdabase{})~\cite{kda2026}.
Both workflows use profiling feedback and candidate history to guide iterative kernel development.
Our experiments cover \numworkloads{} model workloads from KernelBench Level~3~(full machine learning architectures)~\cite{ouyang2025kernelbench} across two accelerator platforms (Ascend~\cite{ascend} and AMD~\cite{amd2025cdna3}) and two language models (GPT-6 Astra~\cite{gpt6astra} and DeepSeek-V4.1-Flash~\cite{deepseekflash}).
We compare \akobase{}+\sysname{} against single-level baselines and alternative cross-level strategies, and use \kdabase{}+\sysname{} to assess reusability across base workflows.
Across four accelerator--model configurations, \akobase{}+\sysname{} improves the equal-weight overall mean speedup by \akoimprovement{} over the strongest baseline by overall mean.
Integrating \sysname{} with \kdabase{} improves \kdabase{}'s arithmetic mean speedup by \kdaimprovement{}.

In this paper, we make the following main contributions:
\begin{itemize}
    \item
    We propose a feedback-guided method that can jointly select source candidates and target abstraction levels for continued kernel development.

    \item
    We design and implement \sysname{} as a pluggable extension of existing agentic kernel workflows.

    \item
    We systematically evaluate \sysname{} across multiple accelerator platforms and language models against single-level baselines and alternative cross-level strategies.
\end{itemize}

\section{Motivation and Problem Definition}
\label{sec:motivation-problem}

\subsection{Motivation}
\label{sec:motivation}

The MinGPT example in Figure~\ref{fig:level-compare} illustrates the different costs and opportunities of development at a fixed abstraction level.
Figure~\ref{fig:level-compare}(a) shows that High-only reaches a strong implementation early, whereas Low-only requires more search to reach higher performance.
Continued high-level development remains limited to the controls exposed by the DSL.
Low-level development exposes additional controls but requires the agent to manage more interacting implementation choices from the beginning.
Exporting a promising high-level candidate allows low-level development to build on the progress already made.
In Figure~\ref{fig:level-compare}(b), the agent continues modifying the exported candidate and obtains further performance improvement.
The exported code makes implementation details previously handled by the compiler available for direct modification.
The agent can refine these details while retaining the computation structure developed during high-level search.

However, cross-level development requires deciding when to change abstraction levels and which candidate to continue from.
A performance plateau may arise because the current representation cannot expose the control needed for further improvement, but the plateau can also appear before the agent finds effective modifications at the current level.
Distinguishing between these possibilities requires examining the candidate's code, performance feedback, and previous modification attempts together.
The agent must identify useful modifications from the available evidence and select a source candidate and target abstraction level that support those modifications.

\subsection{Problem Definition}
\label{sec:problem-definition}

We study how to guide continued kernel development across abstraction levels using existing candidates and evaluation feedback.
The setting consists of a computational task, target hardware $d$, and a base agentic kernel workflow $\mathcal O$.
The base workflow generates, modifies, and evaluates candidate implementations.
We focus on two editable abstraction levels, $\Lambda = \{\mathrm H, \mathrm L\}$, corresponding to high- and low-level representations, respectively.

Correctness checks and measured performance determine candidate quality.
Each candidate $c$ denotes a specific code version of a complete implementation of the task, which may contain multiple kernels.
A candidate is expressed at one abstraction level $\ell(c)\in\Lambda$.
Let $V(c)=1$ indicate that $c$ passes the task's correctness checks.
For a candidate that passes the checks, $m_d(c)$ denotes the execution time metric measured under the evaluation protocol for hardware $d$.
Candidates that have not passed correctness checks may be retained for repair but do not enter performance rankings.

The refinement state records the candidates and feedback available at decision step $t$: $s_t = (c_t, \mathcal C_t, \mathcal H_t)$.
The current candidate is $c_t$, and $\mathcal C_t$ contains the candidates available for further editing, including $c_t$ and retained historical versions.
The history $\mathcal H_t$ records modifications, representation conversions, evaluation results, and available hardware evidence.
Evaluation results and hardware evidence are associated with the corresponding candidate versions and evaluation conditions.

A development decision specifies an action, a source candidate, and a target abstraction level.
Let $\mathcal L(c)\subseteq\Lambda$ denote the target levels available from candidate $c$.
The set includes the current level $\ell(c)$ and lower levels for which conversion is supported.
Returning to a higher level selects a retained high-level source.
We seek a decision rule $\pi$ of the following form where $c_t^{\mathrm{src}}\in\mathcal C_t$ and
$\ell_t^{\mathrm{tar}}\in\mathcal L(c_t^{\mathrm{src}})$.
\begin{equation}
    (a_t, c_t^{\mathrm{src}}, \ell_t^{\mathrm{tar}})
    \leftarrow \pi(s_t)
    \label{eq:level-selection}
\end{equation}

The objective is to improve the performance of the final implementation under a given base workflow and execution protocol.
The execution protocol specifies the run budget, acceptance requirements, and stopping conditions.
The workflow returns the best measured correct candidate as $c_{\mathrm{out}}$, satisfying $V(c_{\mathrm{out}})=1$.
Smaller values of $m_d(c_{\mathrm{out}})$ indicate better performance.

\section{\sysname{}}
\label{sec:method}

\begin{figure}[t]
\centering
\includegraphics[width=0.98\linewidth]{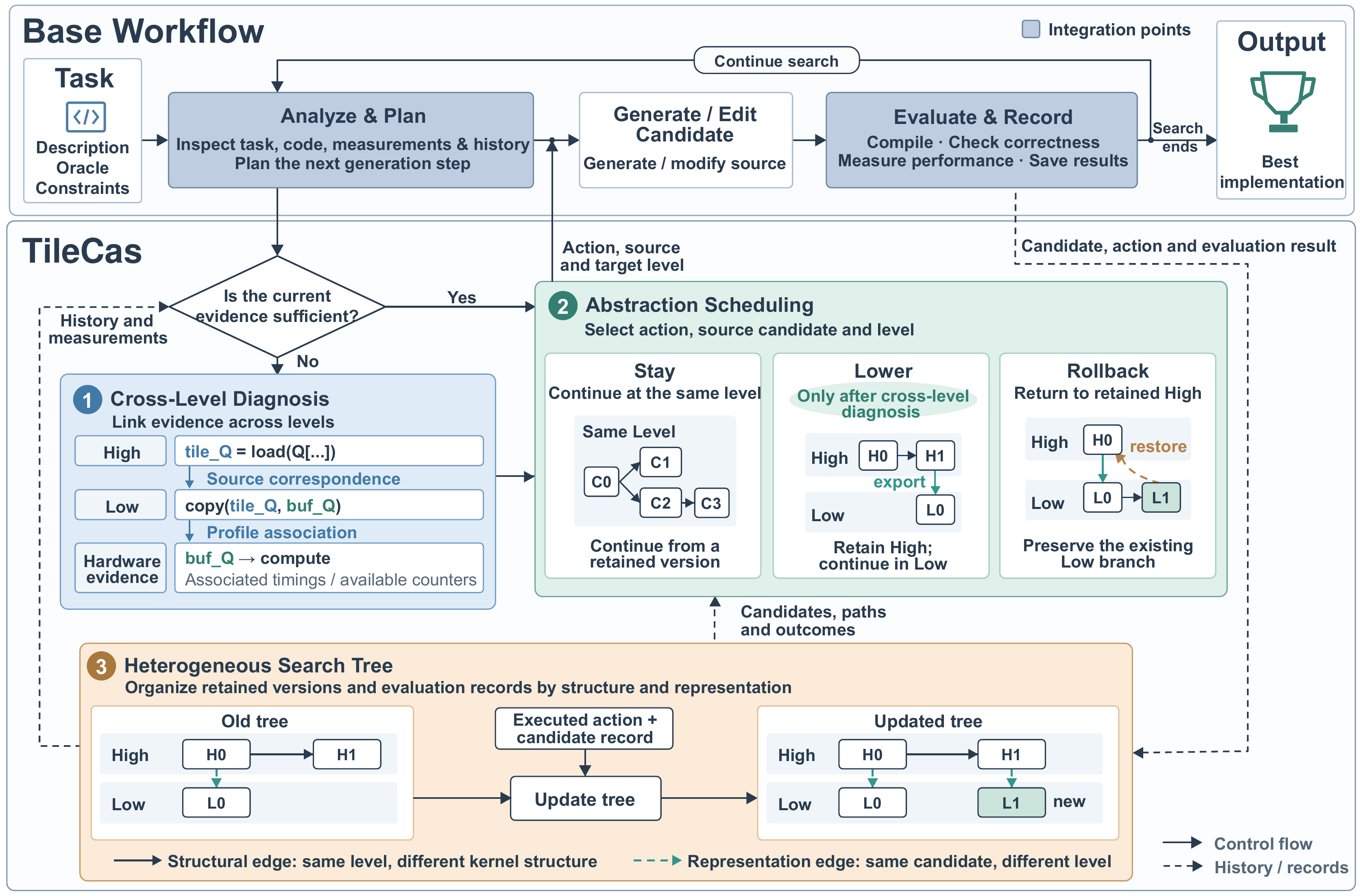}
  \caption{Architecture and workflow of \sysname{}.
  }
  \Description{The upper part shows the base workflow's loop from task input through planning, candidate generation, evaluation, and final output.
  The lower part shows the three components of \sysname{}.
  Planning first checks whether the available evidence is sufficient.
  Insufficient evidence leads to \textit{Cross-Level Diagnosis} before \textit{Abstraction Scheduling}.
  Sufficient evidence leads directly to scheduling.
  Lower requires applicable cross-level diagnostic evidence, which may be reused from earlier iterations.
  Diagnosis links code across abstraction levels with execution evidence, and scheduling returns an action, source candidate, and target level.
  Executed actions, candidates, and evaluation results update the \textit{Heterogeneous Search Tree}.
  The tree records structural derivations and representation conversions and supplies history to both the evidence check and scheduling.}
  \label{fig:overview}
\end{figure}

\subsection{Overview}
\label{sec:overview}

\sysname{} is a unified, pluggable method for feedback-guided programming of accelerator kernels across abstraction levels.
As shown in Figure~\ref{fig:overview}, \sysname{} integrates with the base workflow's \textit{Analyze \& Plan} and \textit{Evaluate \& Record} stages.
Integration requires access to candidate state and evaluation feedback and support for executing the selected plan.
The base workflow retains its generation strategy and evaluation procedure.

The three components connect evidence checks, modification planning, and history updates within each iteration.
The agent first checks whether code, feedback, and history support a concrete next modification.
When evidence is insufficient, \textit{Cross-Level Diagnosis} establishes source correspondence and associates hardware evidence with the executed code.
\textit{Abstraction Scheduling} uses evidence and history to compare controls and implementation requirements across levels and select an action, source candidate, and target level.
After plan execution and evaluation, the \textit{Heterogeneous Search Tree} records candidates, results, and relationships from modifications and conversions, preserving observation context for subsequent decisions.

We use MinGPT causal attention from Figure~\ref{fig:level-compare} as a running example, with TileLang at High and HIP C++ at Low on an AMD Radeon AI PRO R9700.
Figures~\ref{fig:mingpt-correspondence} and~\ref{fig:mingpt-history} illustrate diagnosis and candidate history.

\paragraph{Role of LLMs.}
The base workflow's LLM-based agents follow \sysname{}'s instructions to interpret code and feedback, assess evidence sufficiency, propose modifications, and select actions, source candidates, and target levels.
The coding model generates and revises candidates, and evaluation tools compile the code, check numerical correctness, and measure latency.
Section~\ref{sec:eval-setup} describes the evaluated models and experimental settings.

\subsection{Cross-Level Diagnosis}
\label{sec:cross-level-diagnosis}

\begin{figure}[t]
\centering
\begingroup
\linespread{1}\selectfont

\begin{minipage}[t]{0.58\linewidth}
\vspace{0pt}
{\small\bfseries (a) Source correspondence}
\hfill{\footnotesize\color{twcomment}iter-8}
\par\vspace{3pt}
{\footnotesize\textbf{TileLang}\hfill\texttt{ModelNew.py:54}}
\begin{lstlisting}[style=twcode,language=Python,aboveskip=2pt,belowskip=3pt]
T.reduce_max(s, m, dim=1, clear=False)
\end{lstlisting}
{\footnotesize\textbf{Generated HIP C++}\hfill\texttt{kernel\_4.cpp:92--103}}
\begin{lstlisting}[style=twcode,language=C++,aboveskip=2pt,belowskip=0pt]
for (int i_8 = 0; i_8 < 8; ++i_8) {
  // Local maximum over this row ...
  __syncthreads();
  m_clear[i_8] = tl::AllReduce<tl::MaxOp,
      256, 128, 0>::run(m_clear[i_8],
      (&(((float*)workspace)[0])));
  m_clear[i_8] = tl::AllReduce<tl::MaxOp,
      16, 1, 0>::run(m_clear[i_8]);
  m[i_8] = max(m[i_8], m_clear[i_8]);
}
\end{lstlisting}
\end{minipage}\hfill
\begin{minipage}[t]{0.38\linewidth}
\vspace{0pt}
{\small\bfseries (b) Diagnosis}
\par\vspace{3pt}
{\footnotesize\bfseries Code observations}
\par\vspace{2pt}
{\footnotesize
\renewcommand{\arraystretch}{1.05}
\begin{tabularx}{\linewidth}{@{}l X@{}}
\toprule
Property & Observation \\
\midrule
Memory & Shared memory \\
Synchronization & Block barrier \\
Operation & Max reduction \\
\bottomrule
\end{tabularx}}
\par\vspace{5pt}
{\footnotesize\bfseries Measured performance}
\par\vspace{2pt}
\begingroup
\setlength{\fboxsep}{4pt}
\colorbox{twgray}{%
  \begin{minipage}{\dimexpr\linewidth-2\fboxsep\relax}
    {\footnotesize Whole attention call}
    \par\vspace{2pt}
    {\normalsize\bfseries\color{twblue}2.786430}\enspace{\footnotesize ms}
  \end{minipage}%
}
\endgroup
\par\vspace{3pt}
{\footnotesize\color{twcomment}
The timing covers the entire attention call,
not the reduction alone.}
\end{minipage}
\par
\endgroup
\caption{\textit{Cross-level diagnosis} of MinGPT at iter-8.
(a) A TileLang reduction and a fragment of the associated generated HIP C++ implementation, shown as a code-inspection example.
(b) Code observations and measured whole-attention-call latency for the same candidate.}
\Description{\textit{Cross-level diagnosis} of MinGPT at iter-8.
(a) A TileLang reduction and a fragment of the associated generated HIP C++ implementation, shown as a code-inspection example.
(b) Code observations and measured whole-attention-call latency for the same candidate.}
\label{fig:mingpt-correspondence}
\end{figure}

\textit{Cross-Level Diagnosis} helps the agent interpret execution feedback by examining a candidate's implementation across abstraction levels.
The agent first checks whether the available code, feedback, and history provide sufficient evidence for a concrete next modification.
Evidence is sufficient when available code, feedback, or applicable earlier diagnostic findings allow the agent to justify a modification, identify the code to change, and state an expected outcome to test.
When the evidence is insufficient, the agent invokes \textit{Cross-Level Diagnosis} to establish the relevant connections through \emph{Source correspondence} and \emph{Profile association}.

\emph{Source correspondence} connects a higher-level call to its generated lower-level implementation for a specific candidate version.
During conversion, \sysname{} records the higher-level call site, the generated code, and how the call's arguments are passed to the generated implementation.
\textit{Cross-Level Diagnosis} uses these records to identify the generated implementation associated with the call.
In the running example, the component checks whether the call site, generated code, and argument bindings observed during execution match the conversion records.
The agent inspects the associated code to identify the operations and data dependencies implementing the higher-level call.

\emph{Profile association} links each measurement to the kernel or call group evaluated in a specific candidate.
\textit{Cross-Level Diagnosis} matches kernel identifiers or call relationships in the execution records to the corresponding kernel or call group in the candidate code.
The association retains the candidate version and execution conditions recorded for the measurement.
If the execution records do not identify a unique match, the measurement remains unassociated.
The agent uses the measurement association to locate the measured code and follows source correspondence to inspect the related higher-level calls.
For timing measurements, the recorded time covers the kernel or call group as a whole.
Attributing part of the recorded time to an individual memory access or synchronization operation requires additional evidence.

To check whether earlier evidence remains applicable, \textit{Cross-Level Diagnosis} compares the selected candidate's code and argument bindings with the recorded source correspondence.
For edited lower-level code, the recorded higher-level parent identifies the implementation's origin but does not establish an exact correspondence.
The agent checks whether the relevant code and execution conditions still support the earlier diagnostic findings.

In the running example, the two associations guide inspection of the attention implementation at restored High iter-8.
Attention takes 2.786\,ms, compared with 1.759\,ms for QKV projection and 0.543\,ms for activation conversion.
The attention cost motivates inspection of the associated Low-level code.
Figure~\ref{fig:mingpt-correspondence} shows a High-level softmax reduction alongside a reduction fragment from the associated Low-level implementation.
The call-level association identifies the generated implementation, while the operation-level comparison in the figure comes from code inspection.
The fragment contains shared-memory exchanges and block synchronization, but the recorded time covers the whole attention call.

\begin{figure}[t]
\centering
\includegraphics[width=0.98\textwidth]{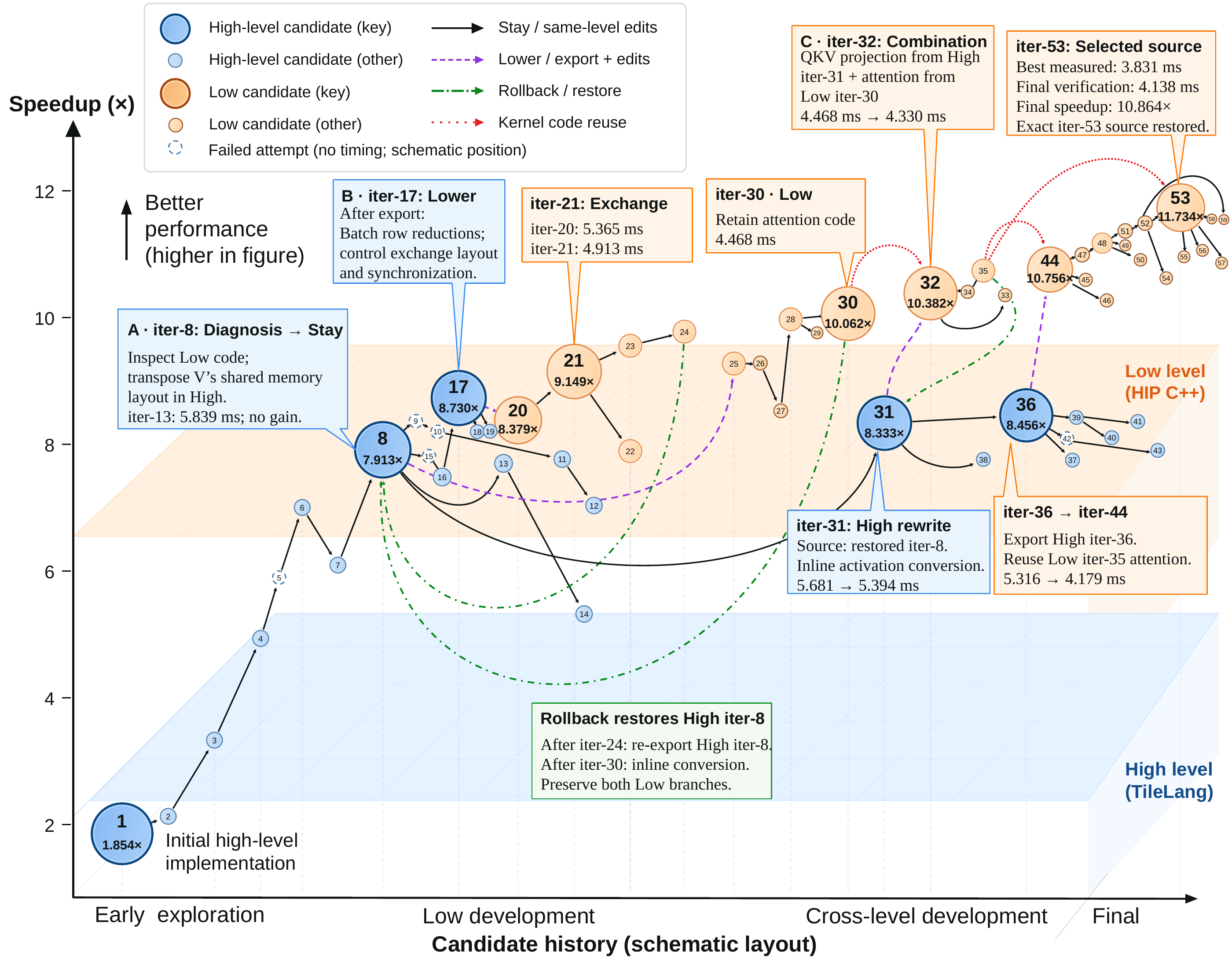}
\caption{
MinGPT candidate history across High (blue) and Low (orange) development.
Callouts \textbf{A}, \textbf{B}, and \textbf{C} highlight Stay, Lower, and cross-branch code reuse, respectively.
}
\Description{
MinGPT candidate history showing same-level development, conversion to Low-level code, restoration of retained High candidates, and reuse of Low-level kernels across branches.
The history retains unsuccessful attempts alongside candidates selected for continued development.
}
\label{fig:mingpt-history}
\end{figure}

A separate observation in the same Low-level attention implementation motivates a hypothesis about the shared-memory layout of $V$.
The generated code uses strided scalar reads for the $PV$ operands, and vector reads for the $QK$ operands.
The difference in access patterns motivates testing whether changing the shared-memory layout of $V$ improves operand access.
Forming the hypothesis from Low-level code does not change the active abstraction level.
Section~\ref{sec:abstraction-scheduling} explains how the agent selects the implementation level for this hypothesis, completing example~\textbf{A} in Figure~\ref{fig:mingpt-history}.

\begin{algorithm}[t]
  \caption{\textit{Abstraction Scheduling} in \sysname{}.}
  \label{alg:abstraction-scheduling}
  \begingroup
  \small
  \linespread{1}\selectfont
  \begin{algorithmic}
    \setlength{\itemsep}{0pt}
    \setlength{\parsep}{0pt}
    \Require Refinement state $s_t=(c_t,\mathcal C_t,\mathcal H_t)$, task and hardware information $\tau$
    \Ensure Plan $p$, or $\bot$
    \State $E\gets\mathrm{ExistingEvidence}(s_t,\tau)$
    \If{\textbf{not} $\mathrm{EvidenceSufficient}(E)$}
      \State $E\gets\mathrm{CrossLevelDiagnosis}(s_t,E)$
    \EndIf
    \If{\textbf{not} $\mathrm{EvidenceSufficient}(E)$}
      \State \Return $\bot$
    \EndIf
    \State $\mathcal D\gets\mathrm{ProposeChanges}(E)$
    \State $\mathcal P\gets\mathrm{FeasiblePlans}(\mathcal D,s_t,E,\tau)$
    \If{$\mathcal P=\varnothing$}
      \State \Return $\bot$
    \EndIf
    \State \Return $\mathrm{SelectPlan}(\mathcal P,E,\mathcal H_t)$
  \end{algorithmic}
  \endgroup
\end{algorithm}

\subsection{Abstraction Scheduling}
\label{sec:abstraction-scheduling}

\textit{Abstraction Scheduling} turns the available evidence into a plan for the next modification, including its source candidate and target abstraction level.
The agent first reviews candidate code, evaluation feedback, and history, invoking \textit{Cross-Level Diagnosis} when the evidence is insufficient under the criterion in Section~\ref{sec:cross-level-diagnosis}.
The agent then forms a testable modification hypothesis and identifies the controls needed to implement the change.

The agent identifies relevant source candidates from the current implementation and summaries of earlier modifications and evaluation results, and reads the relevant code and diagnostic records as needed.
For each source under consideration, the agent checks whether its structure supports the modification and which levels expose the required controls.
The supporting evidence must apply to that source and its execution conditions.
When several choices are feasible, the agent compares their expected outcomes, implementation effort, and validation requirements.
An earlier or slower candidate may be selected when its structure makes the modification easier to implement and evaluate.

Execution restores the selected source when necessary, and then follows one of three actions.
\emph{Lower} exports a supported lower-level representation and requires cross-level diagnostic evidence applicable to the source and proposed modification. 
\emph{Rollback} restores a retained source whose abstraction level is higher than $\ell(c_t)$ and continues development there.
\emph{Stay} covers the remaining cases in which development continues in the source's existing representation, including resuming a retained Low branch.
Existing branches remain available after these actions.

The selected plan $p=(\Delta,a,c^{\mathrm{src}},\ell^{\mathrm{tar}})$ specifies the modification, action, source candidate, and target level, together with a short rationale.
The agent records the selected plan in the candidate history.
Explicit enumeration or ranking of alternative plans is not required.
After execution, the base workflow checks correctness and records the evaluation results in the \textit{Heterogeneous Search Tree} for subsequent decisions.
Only candidates that pass correctness checks enter performance comparisons.
Algorithm~\ref{alg:abstraction-scheduling} summarizes the agent's planning steps.
A return value of $\bot$ indicates that no supported plan is available and leaves the next step to the base workflow's execution protocol.

For the $V$-layout hypothesis in example~\textbf{A} of Figure~\ref{fig:mingpt-history}, TileLang exposes the required shared-array shape, write indices, and matrix-operand orientation.
The agent selects Stay from restored High iter-8, although the hypothesis arises from inspecting Low-level code.
Evaluation confirms correctness but shows no performance improvement.

In example~\textbf{B} of Figure~\ref{fig:mingpt-history}, the agent selects Lower from the retained High candidate iter-17.
Iter-17 remains the fastest correct High candidate after subsequent tile-size and $V$-layout variants fail to improve performance.
After exploring several High structures, the agent selects Low for direct control over cache hints, loop unrolling, and load scheduling.
Inspection of the exported code reveals serial row reductions and motivates batching to reduce repeated synchronization.
The agent tests the change by directly modifying the reduction helpers, shared-memory exchange layout, and synchronization order.
The first attempt is slower, while a subsequent revision improves on the High source.

\subsection{Heterogeneous Search Tree}
\label{sec:heterogeneous-search-tree}

\textit{Heterogeneous Search Tree} preserves candidate implementations and their development histories to support subsequent diagnosis and scheduling.
Each candidate is retained as a complete, versioned implementation linked to an iteration record.
The record identifies the candidate's abstraction level, source parent, and associated evaluation and diagnostic records.
Observations remain associated with the evaluated code version and execution conditions.
The current candidate and the best correct candidate are tracked separately, allowing further exploration while retaining a candidate for final output.

Derivation relationships distinguish same-level modifications from representation conversions.
A \emph{Structural edge} records a structural modification at the same level, while a \emph{Representation edge} records an actual conversion.
Local parameter revisions retain their versions and source relationships within a branch.
Restoring an earlier candidate preserves existing branches, and subsequent modifications create descendants of the restored source.

The agent uses the retained candidates and records to recover suitable starting points and assess whether earlier observations still apply.
Evidence is reused after checking the relevant code and execution conditions.
For cross-branch code reuse, the agent checks the computation and input/output interfaces and records the contributing candidate.
The complete combined candidate undergoes the base workflow's correctness checks and performance evaluation.

Example~\textbf{C} in Figure~\ref{fig:mingpt-history} shows how retained candidates support cross-branch code reuse.
Recorded activation conversion overhead and input-tile reuse in the grouped QKV projection motivate testing inline activation conversion.
The agent restores High iter-8 to modify its grouped QKV projection, producing High iter-31.
The agent selects iter-30, the best retained Low candidate, to reuse the attention improvements developed on that branch.
After exporting iter-31, the agent checks that the attention computation and input/output interfaces remain compatible and reuses the code from iter-30.
The combined candidate iter-32 passes correctness checks and runs in 4.330\,ms, compared with 4.468\,ms for iter-30.

\begin{figure}[t]
\centering
\includegraphics[width=0.7\textwidth]{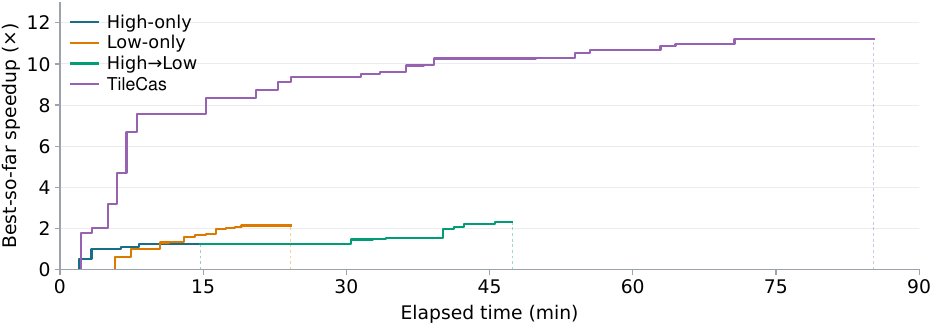}
\caption{
Best-so-far speedup over elapsed time for MinGPT on an AMD Radeon AI PRO R9700.
The comparison retains the High-only, Low-only, and High$\rightarrow$Low trajectories from Figure~\ref{fig:level-compare}(b) and adds the \sysname{}-guided trajectory.
}
\Description{
Performance trajectories for High-only, Low-only, High-to-Low, and \sysname{}, showing best-so-far speedup against elapsed time.}
\label{fig:mingpt-running-trajectory}
\end{figure}

Figure~\ref{fig:mingpt-running-trajectory} extends the MinGPT comparison in Figure~\ref{fig:level-compare}(b) by adding the \sysname{}-guided trajectory.

\section{Evaluation}
\label{sec:evaluation}

To evaluate \sysname{}, we conduct experiments addressing four research questions~(RQs):

\begin{description}
\item[\textbf{RQ1: End-to-End Effectiveness.}]
How effective is \sysname{} in improving end-to-end kernel programming across platforms, coding models, and base workflows?

\item[\textbf{RQ2: Search Efficiency.}]
How does \sysname{} compare with alternative abstraction-level strategies in terms of search efficiency?

\item[\textbf{RQ3: Component Contributions.}]
How do Cross-Level Diagnosis, Abstraction Scheduling, and the Heterogeneous Search Tree respectively affect \sysname{} performance?

\item[\textbf{RQ4: Cross-Level Search Behavior.}]
How effective are the cross-level scheduling decisions made by \sysname{}?
\end{description}

\subsection{Experimental Setup}
\label{sec:eval-setup}

\paragraph{Workloads, platforms and models.}
We evaluate \sysname{} on 50 end-to-end workloads from KernelBench Level~3~\cite{ouyang2025kernelbench}.
Each workload's reference implementation serves as both the correctness reference and performance baseline.
We use the Ascend 910B3~\cite{ascend} and AMD Radeon AI PRO R9700~\cite{amdR9700}, with TileLang~\cite{wang2025tilelang} as the high-level representation and Ascend~C~\cite{huawei_ascendc} and HIP~C++~\cite{amd_hip} as the respective low-level representations.
We use GPT-6 Astra~\cite{gpt6astra} and DeepSeek-V4.1-Flash~\cite{deepseekflash}, yielding four accelerator--model configurations.
Both models use low reasoning effort.
Software details are available on the project website~\cite{projectwebsite}.

\paragraph{Base workflows and configurations.}
We use \akobase{}~\cite{ako2026} as the primary base workflow and evaluate five baselines: High, Low, Fixed, Unrestricted, and Random.
High and Low restrict editing to the corresponding representation.
Fixed performs an independent High-stage search, lowers its best validated candidate once, and starts a fresh Low-stage search without inherited context or history.
Unrestricted retains \akobase{}'s existing multi-representation support without the three \sysname{} components.
Random uses the same legal cross-level actions as \sysname{} but chooses transitions through frequency-matched random routing.
We compare these with \akobase{}+\sysname{}.

We additionally compare \kdabase{}~\cite{kda2026} with \kdabase{}+\sysname{} to evaluate reuse.
Because \kdabase{} does not natively expose the High/Low interface, constructing representation-specific controls would require additional execution paths, conversion, and cross-stage handoff beyond its original workflow.
We therefore compare the original \kdabase{} directly with \kdabase{}+\sysname{}.
RQ1 uses full configuration names; RQ2--RQ4 use short names for \akobase{}-based configurations.

\paragraph{Common evaluation protocol.}
Each method is evaluated on every workload--configuration pair using five independent agent seeds and the same evaluator.
Each run has a 120-minute total budget including failed attempts, diagnosis, representation transitions, and candidate evaluation, although the base workflow may terminate earlier.

Correctness is checked against the KernelBench reference on the same generated inputs with tolerance $10^{-3}$.
Latency is the median of five timed executions after three warm-ups.
A workload--configuration pair is successful if at least three of five runs produce a correct final implementation.
Across pairs, the median and P90 coefficients of variation are 2.20\% and 3.99\%, respectively.

\paragraph{Metrics and aggregation.}
Section~2.2 defines the single-run objective using correctness $V(c)$ and absolute execution time $m_d(c)$: among correct candidates, lower $m_d(c_{\mathrm{out}})$ is better.
For cross-workload evaluation, we normalize the same latency quantity by the workload-specific reference.
For a correct candidate $c$ on hardware $d$, with $c_{\mathrm{ref}}$ denoting the corresponding KernelBench reference, we define
\[
S(c)=\frac{m_d(c_{\mathrm{ref}})}{m_d(c)}.
\]
Thus, $S(c)$ is the speedup of an individual candidate, not an average.
A seed-level run without a correct final implementation contributes zero.
We first average endpoint scores across five seeds for each workload--configuration pair and then across workloads; the overall mean weights the four accelerator--model configurations equally.
Correctness, wall-clock time, and candidate evaluations separately characterize reliability and search cost.

We report standard deviation across five seed-level overall means and compute 95\% confidence intervals using workload-level percentile bootstrap with paired observations preserved.
Seed-level results and resampling details are available online~\cite{projectwebsite}.

For paired comparisons, if neither method meets majority correctness, the result is a tie; if only one does, it wins.
Otherwise, let $S_1$ and $S_2$ be the five-seed mean endpoint speedups of the two methods on the same workload--configuration pair.
We count a tie when
\[
\frac{|S_1-S_2|}{\max(S_1,S_2)}<0.03,
\]
and otherwise assign the win to the larger speedup.

For search-progress analyses, we average each run's best correct speedup at each reporting budget across seeds and workloads.
A run contributes zero until it becomes correct and carries forward its final value after termination.

\subsection{RQ1: End-to-End Effectiveness}
\label{sec:eval-rq1}

We compare \akobase{}+\sysname{} and \kdabase{}+\sysname{} with the five \akobase{} baselines and \kdabase{} on all 200 workload--configuration pairs.
We measure endpoint speedup, majority-success correctness, and paired workload-level outcomes.
Table~\ref{tab:rq1-endpoint} summarizes endpoint performance.
\akobase{}+\sysname{} achieves the highest mean speedup in every
accelerator--model configuration:
4.576$\times$ on Ascend/Astra,
6.722$\times$ on AMD/Astra,
2.778$\times$ on Ascend/DeepSeek Flash, and
5.268$\times$ on AMD/DeepSeek Flash.
Its equal-weight overall mean is 4.836$\times$, compared with
3.748$\times$ for \akobase{}-Random, the strongest \akobase{} baseline by
overall mean, corresponding to a 29.0\% relative improvement.
Among deterministic baselines, \akobase{}-Fixed is strongest at
3.741$\times$, which \akobase{}+\sysname{} improves by 29.3\%.
The overall 95\% intervals are $[4.054,5.704]$ for
\akobase{}+\sysname{} and $[3.140,4.434]$ for
\akobase{}-Random.

The gain is not driven by a single accelerator--model configuration.
Relative to the strongest \akobase{} baseline within each configuration,
\akobase{}+\sysname{} improves mean speedup by
22.0\% on Ascend/Astra,
22.5\% on AMD/Astra,
21.2\% on Ascend/DeepSeek Flash, and
13.7\% on AMD/DeepSeek Flash.
Random is strongest on the two Astra configurations, while Fixed is strongest
on the two DeepSeek Flash configurations.

\paragraph{Stability across agent runs.}
The endpoint results remain stable across the five independent agent seeds.
At the aggregate level, the standard deviation is
0.221$\times$ for \akobase{}+\sysname{} and
0.167$\times$ for \akobase{}-Random, the strongest overall baseline.
For the second base workflow, the corresponding values are
0.119$\times$ for \kdabase{}+\sysname{} and
0.044$\times$ for \kdabase{}.
Together with the 2.20\% median and 3.99\% P90 coefficient of variation
reported in the setup, these results indicate low sensitivity to agent
stochasticity relative to the observed endpoint improvements.

\paragraph{Per-workload behavior.}
The arithmetic-mean improvements are also broad across workloads.
Against \akobase{}-Random, \akobase{}+\sysname{} records
158 wins, 5 ties, and 37 losses under the 3\% relative tie threshold,
corresponding to a 79.0\% win rate.
Against \akobase{}-Fixed, it records 144 wins, 6 ties, and 50 losses,
corresponding to a 72.0\% win rate.
Against \kdabase{}, \kdabase{}+\sysname{} records
159 wins, 3 ties, and 38 losses, corresponding to a 79.5\% win rate.

\begin{table*}[t]
\centering
\small
\caption{End-to-end endpoint performance.
Overall mean is reported with standard deviation across five independent agent seeds; correctness uses majority success. Best values are bolded.}
\label{tab:rq1-endpoint}
\scriptsize
\setlength{\tabcolsep}{3.5pt}
\begin{tabular}{@{}lccccccc@{}}
\toprule
Method &
Asc./Astra &
AMD/Astra &
Asc./DS &
AMD/DS &
Overall mean &
Overall median &
Correct \\
\midrule
\akobase{}-High &
2.402 & 5.013 & 1.046 & 3.792 &
3.063$\pm$0.082 & 2.301 & 188/200 \\

\akobase{}-Low &
3.032 & 3.153 & 1.981 & 2.915 &
2.770$\pm$0.085 & 2.208 & 192/200 \\

\akobase{}-Fixed &
2.982 & 5.055 & 2.293 & 4.632 &
3.741$\pm$0.113 & 3.037 & 188/200 \\

\akobase{}-Unrestricted &
2.444 & 4.696 & 1.856 & 3.942 &
3.235$\pm$0.123 & 2.473 & 192/200 \\

\akobase{}-Random &
3.750 & 5.487 & 1.962 & 3.792 &
3.748$\pm$0.167 & 2.804 & 191/200 \\

\kdabase{} &
2.429 & 3.068 & 2.007 & 2.956 &
2.615$\pm$0.044 & 2.131 & 190/200 \\

\akobase{}+\sysname{} &
\textbf{4.576} & \textbf{6.722} &
\textbf{2.778} & \textbf{5.268} &
\textbf{4.836$\pm$0.221} &
\textbf{3.767} & \textbf{195/200} \\

\kdabase{}+\sysname{} &
4.437 & 6.339 & 2.371 & 4.553 &
4.425$\pm$0.119 & 3.392 & \textbf{195/200} \\
\bottomrule
\end{tabular}
\end{table*}

\begin{table}[t]
\centering
\small
\caption{Paired workload-level endpoint comparisons in RQ1.
Each comparison contains the 200 workload--configuration pairs and uses the
five-seed aggregated endpoint.}
\label{tab:rq1-paired}
\small
\begin{tabular}{@{}lcc@{}}
\toprule
Comparison & Win / Tie / Loss & Win rate \\
\midrule
\akobase{}+\sysname{} vs.\ \akobase{}-Random &
158 / 5 / 37 & 79.0\% \\
\akobase{}+\sysname{} vs.\ \akobase{}-Fixed &
144 / 6 / 50 & 72.0\% \\
\kdabase{}+\sysname{} vs.\ \kdabase{} &
159 / 3 / 38 & 79.5\% \\
\bottomrule
\end{tabular}
\end{table}

\paragraph{Reuse across base workflows.}
Relative to \akobase{}-Unrestricted, adding \sysname{} increases the
overall mean from 3.235$\times$ to 4.836$\times$, a 49.5\% improvement.
For \kdabase{}, the same interface increases the overall mean from
2.615$\times$ to 4.425$\times$, a 69.2\% improvement.
The \kdabase{} integration improves mean speedup in all four
accelerator--model configurations.

\paragraph{Discussion.}
\akobase{}-Random is the strongest baseline by overall mean, although
\akobase{}-Fixed is within 0.2\% of it.
The strongest baseline also varies by configuration: Random is strongest on
the two Astra configurations, while Fixed is strongest on the two DeepSeek
Flash configurations.
This indicates that no single abstraction-level strategy is universally best.
Correctness failures are also more frequent under DeepSeek-V4.1-Flash than
under Astra.
Because failed seed-level runs contribute zero, this reliability difference
is already reflected in endpoint speedup.

\noindent\textbf{Key finding.}
\akobase{}+\sysname{} achieves the highest mean speedup in all four
accelerator--model configurations and improves the strongest \akobase{}
baseline by 29.0\%.
The gain is stable across five independent agent seeds and broad across
individual workloads.
The same interface improves \kdabase{} by 69.2\%.

\subsection{RQ2: Search Efficiency}
\label{sec:eval-rq2}

\begin{figure*}[t]
    \centering
    \includegraphics[width=0.75\textwidth]{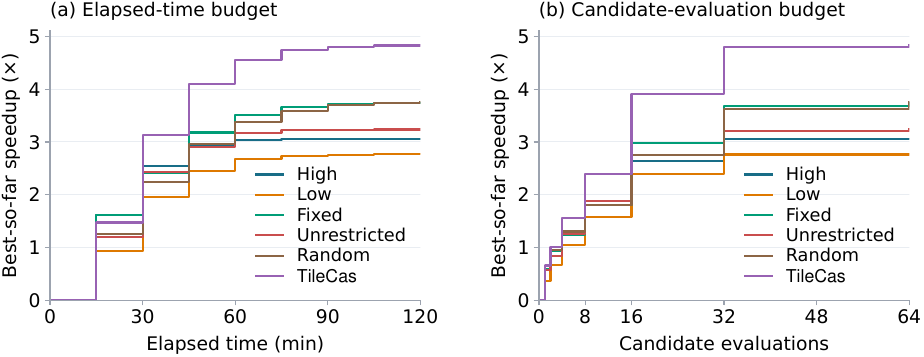}
    \caption{Search progress.
    Best-so-far speedup is shown under
    (a) elapsed wall-clock time and
    (b) candidate-evaluation budget.}
    \Description{Two best-so-far search curves comparing High, Low, Fixed,
    Unrestricted, Random, and TileCas under elapsed-time and
    candidate-evaluation budgets.}
    \label{fig:search-progress}
\end{figure*}

We compare High, Low, Fixed, Unrestricted, Random, and \sysname{} under the same search budget. 
We measure both the cost of reaching the first correct implementation and best-so-far speedup as a function of elapsed wall-clock time and candidate evaluations.
As shown in Table~\ref{tab:rq2-first-correct}, High reaches the first correct candidate fastest:
6.81 minutes and 1.43 candidate evaluations on average.
Random requires 10.58 minutes and 1.85 evaluations, while
\sysname{} requires 10.74 minutes and 1.80 evaluations and Low requires
12.17 minutes and 2.06 evaluations.
Thus, \sysname{}'s endpoint advantage does not arise from producing the first
correct implementation sooner.

\begin{table}[t]
\centering
\caption{Cost of reaching the first correct implementation.
Successful counts use majority success across five agent seeds.}
\label{tab:rq2-first-correct}
\small
\begin{tabular}{@{}lccc@{}}
\toprule
Method & Successful & Time (min)$\downarrow$ &
Evaluations$\downarrow$ \\
\midrule
High & 188/200 & \textbf{6.81} & \textbf{1.43} \\
Low & 192/200 & 12.17 & 2.06 \\
Fixed & 188/200 & 11.39 & 1.82 \\
Unrestricted & 192/200 & 11.96 & 1.94 \\
Random & 191/200 & 10.58 & 1.85 \\
\sysname{} & \textbf{195/200} & 10.74 & 1.80 \\
\bottomrule
\end{tabular}
\end{table}

The ordering changes during continued refinement.
At 30 minutes, \sysname{} reaches 3.134$\times$, compared with
2.543$\times$ for High, 2.403$\times$ for Fixed, and
2.238$\times$ for Random.
At 60 minutes, it reaches 4.553$\times$, compared with
3.503$\times$ for Fixed and 3.376$\times$ for Random.
By 90 minutes, \sysname{} reaches 4.806$\times$, while Fixed and Random
reach 3.720$\times$ and 3.695$\times$, respectively.

The same pattern remains when budget is measured by candidate evaluations.
At 16 evaluations, \sysname{} reaches 3.904$\times$, compared with
2.975$\times$ for Fixed, 2.744$\times$ for Random, and
2.642$\times$ for High.
At 32 evaluations, \sysname{} reaches 4.790$\times$, compared with
3.679$\times$ for Fixed, 3.628$\times$ for Random, and
3.049$\times$ for High.

\paragraph{Search saturation.}
The trajectories exhibit clear saturation near the end of the search budget.
For \sysname{}, mean best-so-far speedup increases only from
4.806$\times$ at 90 minutes to 4.836$\times$ at 120 minutes,
an additional improvement of approximately 0.6\%.
Likewise, doubling the candidate-evaluation budget from 32 to 64 increases
mean speedup only from 4.790$\times$ to 4.836$\times$,
an additional improvement of approximately 1.0\%.
The baselines show the same qualitative flattening, indicating that the
workflows approach search saturation within the allocated budget.

\paragraph{Discussion.}
The faster first correctness of High does not imply more effective search.
A key feature of \sysname{} is that exploratory failures are recoverable:
Cross-Level Diagnosis turns unsuccessful attempts into actionable feedback,
the Heterogeneous Search Tree preserves validated states and alternative
branches, and Abstraction Scheduling can redirect development when an attempt
does not succeed.
Together, these mechanisms allow less conservative exploration without
requiring every intermediate candidate to be immediately correct.

An Ascend/Astra MLP run illustrates this behavior.
The High-stage baseline reaches correctness after 4.13 minutes, whereas
\sysname{} requires 5.41 minutes because its first, more aggressive
High-level candidate fails compilation and is repaired in the next iteration.
Both attempts occur before any Lower or Rollback, so the delay is not caused
by cross-level transition overhead.
The complete trace is available online~\cite{projectwebsite}.

The advantage of \sysname{} also persists when progress is measured by
candidate evaluations rather than elapsed time, making a purely
per-iteration runtime explanation unlikely.
The saturation under both budget views further shows that \sysname{} reaches
a stronger performance level rather than merely postponing improvements to
the end of a longer search.

\noindent\textbf{Key finding.}
High reaches correctness sooner, but \sysname{} uses the subsequent search
budget more effectively.
Its advantage persists under both wall-clock and candidate-evaluation budgets
and reaches a substantially stronger saturation level.

\subsection{RQ3: Component Contributions}
\label{sec:eval-rq3}

We perform one-at-a-time ablations of the three \sysname{} components while keeping the common workload, model, evaluator, seed, and search-budget protocol unchanged.
w/o Diagnosis removes Cross-Level Diagnosis.
w/o Scheduling replaces evidence-guided Abstraction Scheduling with the fixed
High-to-Low progression.
w/o Tree removes editable access to alternative historical branches while
retaining the incumbent.

\paragraph{Results.}
Table~\ref{tab:rq3-ablation} shows that every ablation reduces endpoint
performance.
Removing Diagnosis lowers mean speedup from 4.836$\times$ to
4.229$\times$, a 12.6\% reduction.
Removing Scheduling lowers it to 3.854$\times$, the largest reduction at
20.3\%.
Removing the Heterogeneous Search Tree lowers it to
4.141$\times$, a 14.4\% reduction.

\begin{table*}[t]
\centering
\small
\caption{
Ablation results.
Mean speedup is reported with standard deviation across five agent seeds.
Correctness uses majority success.
Loss is relative to the complete \sysname{} configuration.}
\label{tab:rq3-ablation}
\small
\begin{tabular}{@{}lccccc@{}}
\toprule
Variant &
Correct (\%)$\uparrow$ &
Mean speedup$\uparrow$ &
Evaluations$\downarrow$ &
Wall time (min)$\downarrow$ &
Loss \\
\midrule
Full \sysname{} &
\textbf{97.5} &
\textbf{4.836$\pm$0.221} &
21.76 & 52.05 & -- \\

w/o Diagnosis &
96.5 &
4.229$\pm$0.032 &
22.82 & \textbf{50.14} & 12.6\% \\

w/o Scheduling &
94.5 &
3.854$\pm$0.147 &
22.47 & 53.04 & 20.3\% \\

w/o Tree &
96.5 &
4.141$\pm$0.041 &
22.20 & 54.71 & 14.4\% \\
\bottomrule
\end{tabular}
\end{table*}

The variants consume similar search budgets.
The complete configuration averages 21.76 candidate evaluations and
52.05 minutes, compared with 22.82/50.14, 22.47/53.04, and 22.20/54.71
for the three ablations.
The endpoint differences therefore cannot be explained by the complete
configuration simply evaluating substantially more candidates.

\paragraph{Discussion.}
The ablation losses are not additive.
Diagnosis changes the evidence available to Scheduling, Scheduling changes
which branches are explored, and the Heterogeneous Search Tree changes which
historical states remain available to both mechanisms.
The slightly shorter wall time of w/o Diagnosis is therefore not evidence
that Diagnosis is unnecessary: removing diagnostic work reduces overhead but
also produces a substantially weaker endpoint.

\noindent\textbf{Key finding.}
Removing Diagnosis, Scheduling, or the Heterogeneous Search Tree reduces mean
speedup by 12.6\%, 20.3\%, and 14.4\%, respectively, under comparable search
budgets.

\subsection{RQ4: Cross-Level Search Behavior}
\label{sec:eval-rq4}

RQ4 combines whole-run routing controls, post-action analysis, and matched
interventions to distinguish the value of cross-level movement from the value
of selecting when to move.

\paragraph{Frequency-matched random control.}
For each accelerator--model configuration, we derive the empirical conditional transition rates from the corresponding \sysname{} trajectories:
\[
p_L=\frac{N_{\mathrm{Lower}}}{N_{\mathrm{High\ boundaries}}},
\qquad
p_R=\frac{N_{\mathrm{Rollback}}}{N_{\mathrm{Low\ boundaries}}}.
\]
At a High boundary, Random selects Lower with probability $p_L$ and Stay otherwise.
At a Low boundary, Random selects Rollback with probability $p_R$ and Stay otherwise.
The frequency-matched Random policy uses aggregate conditional probabilities
of 14.22\% for Lower given High and 5.88\% for Rollback given Low.
Random otherwise follows the common evaluation protocol and ignores diagnostic
evidence when choosing when to transition.

\paragraph{Whole-run routing results.}
Table~\ref{tab:rq4-policies} compares the complete routing policies.
Random reaches 3.748$\times$ mean speedup, whereas \sysname{} reaches
4.836$\times$, a 29.0\% relative improvement.
The difference holds in all four accelerator--model configurations:
22.0\% on Ascend/Astra,
22.5\% on AMD/Astra,
41.6\% on Ascend/DeepSeek Flash, and
38.9\% on AMD/DeepSeek Flash.

\begin{table*}[t]
\centering
\small
\caption{
Whole-run routing-policy comparison.
Mean speedup is reported with standard deviation across five independent
agent seeds.}
\label{tab:rq4-policies}
\small
\begin{tabular}{@{}lccccc@{}}
\toprule
Policy &
Mean speedup$\uparrow$ &
Evals./run$\downarrow$ &
Time (min)$\downarrow$ &
Lower/run &
Rollback/run \\
\midrule
Fixed &
3.741$\pm$0.113 &
23.94 & 54.34 & 0.935 & 0.000 \\

Random &
3.748$\pm$0.167 &
23.28 & 54.71 & 1.368 & 0.723 \\

Unrestricted &
3.235$\pm$0.123 &
\textbf{18.34} & \textbf{40.17} & -- & -- \\

\sysname{} &
\textbf{4.836$\pm$0.221} &
21.76 & 52.05 & 1.255 & 0.540 \\
\bottomrule
\end{tabular}
\end{table*}

Random performs more cross-level movements per run than \sysname{}:
1.368 versus 1.255 Lower actions and
0.723 versus 0.540 Rollbacks.
Its weaker endpoint therefore cannot be explained by receiving fewer
opportunities to move between abstraction levels.

\paragraph{Post-action outcomes.}
A short window is productive if its best correct candidate improves the
pre-action incumbent by more than 3\% within the next three candidate
evaluations.

\begin{table}[t]
\centering
\small
\caption{Performance following routing actions in \sysname{} trajectories.
Productive denotes an improvement of more than 3\% within the next three
candidate evaluations.}
\label{tab:rq4-actions}
\small
\begin{tabular}{@{}lcc@{}}
\toprule
Action & Productive$\uparrow$ & Median gain$\uparrow$ \\
\midrule
Stay & 83.4\% & 13.12\% \\
Lower & \textbf{84.4\%} & \textbf{27.37\%} \\
Rollback & 64.4\% & 6.99\% \\
\bottomrule
\end{tabular}
\end{table}

Lower has the strongest immediate effect:
84.4\% of eligible windows are productive, with a median gain of 27.37\%.
Rollback is weaker in the immediate three-evaluation window.
When the observation window is extended from a Rollback until the next
Rollback or run termination, however, 79.8\% of eligible windows are
productive, with a median gain of 50.07\%.
Because this longer window contains subsequent modifications, the latter
statistic is descriptive rather than an isolated causal effect of Rollback.

\paragraph{Matched interventions.}
To separate action choice from the state at which the action occurs, we sample
25 eligible High switching checkpoints and 25 eligible Low switching
checkpoints.
At High checkpoints, Lower is compared with Stay; at Low checkpoints,
Rollback is compared with Stay.
Both continuations begin from the same candidate state and receive the same
model, evaluator, and local budget of 10 candidate evaluations.

For a continuation, we define
\[
g =
100
\left(
\frac{S_{\mathrm{after}}}
     {S_{\mathrm{before}}}
-1
\right),
\]
where $S_{\mathrm{before}}$ is the checkpoint speedup and
$S_{\mathrm{after}}$ is the best correct speedup obtained within the next
10 candidate evaluations.

At High checkpoints, Lower achieves the better tested continuation in
23 of 25 cases (92.0\%).
At Low checkpoints, Rollback is better in 19 of 25 cases (76.0\%).
Across all 50 interventions, the selected cross-level action is better in
42 cases (84.0\%), with mean regret of 1.92 percentage points and median
regret of zero.

\begin{figure*}[t]
    \centering
    \includegraphics[width=0.8\textwidth]
    {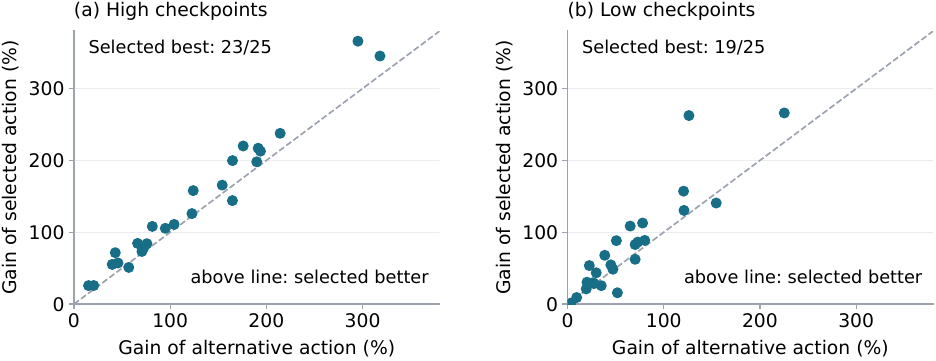}
    \caption{Matched-checkpoint interventions
    in the \sysname{} workflow.
    Each point compares the selected cross-level continuation with Stay from
    the same candidate state under the same 10-evaluation local budget.
    Points above the diagonal favor the selected cross-level action.}
    \Description{Scatter plots comparing selected Lower versus Stay from High
    checkpoints and selected Rollback versus Stay from Low checkpoints.}
    \label{fig:matched-checkpoints}
\end{figure*}

\paragraph{Discussion.}
The whole-run controls progressively eliminate simpler explanations for the
endpoint improvement.
Fixed shows that a predetermined one-way High-to-Low progression is
insufficient.
Random shows that merely making cross-level transitions at similar
frequencies is also insufficient.
The matched interventions instead show that the current search state matters:
the \sysname{}-selected cross-level continuation is the better tested choice
in 84.0\% of sampled switching checkpoints.

Lower has a strong immediate effect, whereas Rollback often has delayed
value.
Rollback can leave a low-value local direction and reopen a retained High
branch from which later Low candidates follow a different development path.
The complete traces are available online~\cite{projectwebsite}.

The conditional probabilities used by Random are estimated from the same
evaluation cohort.
Random should therefore be interpreted as a post-hoc transition-frequency
control rather than an independently trained or out-of-sample routing policy.

\noindent\textbf{Key finding.}
Cross-level movement is useful because \sysname{} selects when and where to
move, not merely because transitions are available.
\sysname{} outperforms frequency-matched random routing by 29.0\%, and its selected
cross-level action is the better tested continuation in 84.0\% of matched
checkpoints.

\section{Related Work}
\label{sec:related}

\paragraph{Agentic kernel generation and optimization.}
AKO~\cite{ako2026} provides autonomous coding agents with benchmarking, profiling, trajectory recording, and language switching; KDA~\cite{kda2026} structures kernel development around workload contracts, candidate generation, correctness checks, and performance evaluation.
\sysname{} extends these workflows and inherits their optimization capabilities.
Our integrations preserve this development-and-evaluation loop; selecting the programming level is an additional decision made using the available evidence and search history.
Related systems strengthen planning, profiling, and reuse of historical candidates across platforms~\citep{zhu2025qimeng,du2025akgkernelagent,luo2026agentic,zhang2025accelopt,sereda2026kforge,cheng2026kernelagent,atrex2026,dong2026stark,gai2026kernelpro}, while execution-based rewards improve trained kernel generators~\citep{li2025cudal1,su2025cudal2,li2025autotriton}.
EGG~\citep{han2026egg} separates algorithmic design from hardware tuning, and HIERA~\citep{wang2026hiera} selects among framework operators, libraries, and custom kernels.
\sysname{} jointly selects a retained candidate and editable level, supporting continued edits, lowering, and revisiting higher-level branches.
Existing benchmarks broaden workload, platform, generalization, and verification coverage~\citep{ouyang2025kernelbench,wen2025multikernelbench,younesian2026agentkernelarena,atrex2026,nie2026kernelcraft,lange2025robustkbench}; we use KernelBench Level~3~\citep{ouyang2025kernelbench} as the common workload contract.

\paragraph{Agent--compiler cooperation and agent-oriented abstractions.}
ACCLAIM~\cite{mikek2026compilerllm} optimizes C programs on CPUs using correctness and runtime feedback to coordinate LLM rewrites with unchanged compiler components.
The reported feedback does not explicitly associate runtime costs with corresponding code across abstraction levels.
Under an end-to-end model compilation view, \sysname{} augments base agentic workflows as backend components for kernel optimization through cross-level diagnosis.
This diagnosis links measured kernel costs to DSL operations and generated Low-level implementations, helping the agent determine whether a proposed optimization requires DSL-level changes or direct Low-level edits.
Ave~\citep{mai2026argus} (formerly ARGUS) checks data-flow invariants in a tile-based DSL and returns localized counterexamples, supporting targeted repair while hiding lower-level representations.
CAKE~\citep{ye2026cake} exposes warp roles, memory movement, synchronization, and pipelines through a typed IR with verification and localized diagnostics; its IR and compiler evolve with kernel requirements.
Ave and CAKE improve the controls and feedback within their designed abstractions; \sysname{} asks when development should stay at one level or resume elsewhere.
Related approaches~\citep{hari2026mucutlass,gao2026zomboss,shypula2023pie,cummins2023llmopt,cummins2024llmcompiler,tang2025reasoningcompiler} improve agent-facing abstractions, source-level edits, compiler representations, and transformation search.
Optimization environments and learned compiler policies expose further decision spaces~\citep{huang2020autophase,trofin2021mlgo,cummins2021compilergym,liang2023rlcompopt}, while execution feedback and self-refinement improve program selection and repair~\citep{chen2023selfdebug,le2022coderl,ni2023lever,chen2022codet,madaan2023selfrefine,shinn2023reflexion,gou2023critic}.
\sysname{} builds on these ideas through cross-representation evidence association and joint candidate/level selection, preserving alternative branches across existing High and Low representations without requiring a new IR.

\paragraph{Programming abstractions and multi-level search.}
Feedback-guided implementation search predates LLMs: SPIRAL~\citep{puschel2005spiral} optimizes SPL formulas and generated C/Fortran code, while ATLAS~\citep{whaley1998atlas} and FFTW~\citep{frigo2005fftw} adapt numerical implementations to hardware.
Structured languages, compiler infrastructures, and user-extensible schedules expose complementary control over computation and implementation~\citep{ragankelley2013halide,vasilache2018tc,kjolstad2017taco,tillet2019triton,wang2025tilelang,lattner2021mlir,chen2018tvm,ikarashi2022exo,ikarashi2024exo2}.
Autotuning and equality saturation explore or retain implementation alternatives~\citep{chen2018autotvm,zheng2020ansor,ansel2014opentuner,willsey2020egg}; STOKE~\citep{schkufza2013stoke}, BOLT~\citep{panchenko2019bolt}, and CuAsmRL~\citep{he2025cuasmrl} continue optimization after compilation.
Agent pipelines also combine DSL-guided lowering or semantic refactoring with subsequent optimization~\citep{wen2026ascendcraft,qu2026twostage}.
Collectively, these lines of work show the value of structured representations, staged refinement, and search over alternative implementations.
\sysname{} studies how runtime evidence and candidate history guide a coding agent's next programming context: which implementation to resume, which level exposes the needed controls, and which cross-level branches to retain for later continuation.

\section{Limitations and Threats to Validity}
\label{sec:limitations}

\paragraph{Evaluation scope.}
Our evaluation covers two editable abstraction levels, 50 KernelBench Level~3 workloads, two accelerator platforms, and two coding models.
Generalization to additional abstraction levels and compiler representations remains to be evaluated.
A compiler update may expose previously unavailable controls at the high level, changing which level supports a proposed modification.
The selected workloads also limit conclusions about dynamic shapes, communication, and multi-device execution.

\paragraph{Diagnosis and scheduling.}
The quality of diagnosis and scheduling depends on the available evidence and the coding model's interpretation.
Kernel or call-group timings identify expensive regions but provide limited evidence about individual memory accesses or synchronization operations.
An incorrect interpretation can lead to ineffective modifications or unnecessary transitions, consuming the search budget.
Edits and conversions can change the relevance of earlier observations despite preserved candidate histories.

\paragraph{Measurement and variability.}
Correctness and performance results depend on the tested inputs and run-to-run variability.
The output checks use generated inputs and a numerical tolerance of $10^{-3}$, leaving behavior on other inputs unverified.
Warm-up executions and median timing reduce measurement noise, but device state may still affect measured latency.
Model sampling also changes search trajectories.
We quantify this variation using five independent agent seeds and report across-seed variability, while workload-level bootstrap intervals capture uncertainty across workloads.

\paragraph{Intervention scope.}
The matched interventions compare the selected cross-level action against Stay at 50 sampled switching checkpoints under the same 10-evaluation budget.
The results assess local continuation choices and do not evaluate whether Stay decisions at other checkpoints are appropriate.
Longer continuations or different modifications may change which action performs best.
In the separate trajectory analysis, longer Rollback observation windows capture the combined effects of candidate restoration and subsequent edits, so the observed gains cannot be attributed to restoration alone.

\paragraph{Portability.}
Reuse across base workflows has been evaluated through the AKO and KDA integrations.
Additional workflows require interfaces for exposing candidate state and evaluation feedback, executing selected plans, and restoring retained candidates.
New backends also require support for code conversion and source correspondence, and profiler availability may limit the evidence available for diagnosis.
The experiments assess kernel performance after integration and do not quantify the engineering effort required for these adaptations.

\section{Conclusion}
\label{sec:conclusion}

Abstraction levels shape the search space available to coding agents during accelerator-kernel development.
High-level representations simplify structural exploration, while native code exposes finer control over hardware operations.
Restricting development to one level can limit progress, but using both requires deciding what to modify, where to make the change, and which candidate to continue from.

In this paper, we have presented \sysname{} to address these decisions through Cross-Level Diagnosis, Abstraction Scheduling, and a Heterogeneous Search Tree.
These components relate execution feedback to candidate code, guide the next modification and its implementation level, and preserve candidate histories for subsequent decisions.
Evaluation with AKO and KDA demonstrates the benefits of feedback-guided cross-level programming across different base workflows.
The experiments cover 50 KernelBench Level~3 workloads across two accelerator platforms and two language models.
AKO with \sysname{} improves overall arithmetic mean speedup by \akoimprovement{} over the strongest evaluated baseline.
Integrating \sysname{} with KDA improves the same metric by 69.2\% over KDA.
These results support treating abstraction selection as an explicit part of agent-driven kernel development, guided by execution evidence and the history of prior exploration.

\section*{Acknowledgments}
This work was partially supported by the Fundamental and Interdisciplinary Disciplines Breakthrough Plan of the Ministry of Education of China (No. JYB2025XDXM118) and the National Natural Science Foundation of China under Grant Nos. U25A6023, 92464301, and 623B2006.

Tao Xie is also affiliated with the Key Laboratory of High Confidence Software Technologies (Peking University), Ministry of Education, China; Shanghai Institute of Systems for Open Computing, China; and School of Computer Science and Engineering, Southeast University, China.

\section*{Data Availability}
The source code, additional experimental details and results,
and supplementary materials are available on the project website~\cite{projectwebsite}.

\begingroup
\small
\setlength{\bibsep}{4pt}
\bibliography{references}
\bibliographystyle{plainnat}
\endgroup

\end{document}